\documentclass[nohyper,12pt,letterpaper]{JHEP}
\usepackage{epsfig}

\newfont{\frak}{eufm10 scaled 1200}

\newfont{\Bbb}{msbm10 scaled 1200}     
\newcommand{\mathbb}[1]{\mbox{\Bbb #1}}
\DeclareSymbolFont{AMSa}{U}{msa}{m}{n}
\DeclareSymbolFont{AMSb}{U}{msb}{m}{n}
\let\Box\relax
\DeclareMathSymbol{\Box}{\mathord}{AMSa}{"03}

\def \eqn#1#2{\begin{equation}#2\label{#1}\end{equation}}

\title{On Anthropic Solutions of the Cosmological Constant Problem}

\author{T. Banks\thanks{On leave from Rutgers University.}, M. Dine\\
  Department of Physics and Institute for Particle Physics\\
  University of California, Santa Cruz, CA 95064\\
E-mail: \email{banks@scipp.ucsc.edu, dine@scipp.ucsc.edu}}

\author{L. Motl\thanks{This work was done while visiting
UCSC.}\\ Department of Physics and Astronomy\\ Rutgers University,
Piscataway, NJ 08855-0849\\ E-mail:
\email{motl@physics.rutgers.edu}}

\abstract{Motivated by recent work of Bousso and Polchinski
(BP), we study theories which explain the small value of
the cosmological constant using the anthropic principle.  We argue
that simultaneous solution of the gauge hierarchy problem is a
strong constraint on any such theory.  We exhibit three classes of
models which satisfy these constraints.  The first is a version of
the BP model with precisely two large dimensions.  The second
involves 6-branes and antibranes wrapped on supersymmetric
3-cycles of Calabi-Yau manifolds, and the third is a version of the
irrational axion model. All of them have possible
problems in explaining the size of microwave background
fluctuations.    We also find that
most models of this type predict that all constants in the low
energy Lagrangian, as well as the gauge groups and representation
content, are chosen from an ensemble and cannot be uniquely determined
from the fundamental theory.  In our opinion, this significantly reduces
the appeal of this kind of solution of the cosmological constant problem.
On the other hand, we argue that the vacuum selection problem
of string theory might plausibly have an anthropic, cosmological solution.}


\received{???????? ?st, 2000} \accepted{???????? ?th, 1998}

\preprint{\hepth{0007206}\\RUNHETC-2000-23\\SCIPP-00/21}
\begin{document}


\section{Introduction}

In a recent paper \cite{bp}, Bousso and Polchinski (BP) have
revisited and improved some old ideas for explaining the value of
the cosmological constant.  They focused on the membrane creation
scenario of Brown and Teitelboim \cite{brt,brtwo} and found what
they claim to be a consistent anthropic solution of the
cosmological constant problem.  They exhibited models which
potentially have a large number
of discrete vacuum states, a small fraction of which have a small
cosmological constant.  They then invoke Weinberg's bound on
values of the cosmological constant consistent with galaxy
formation \cite{wone} and/or recent improvements on that bound
\cite{gvil}-\cite{wbrecent} (see however \cite{wbdist})
to argue that only those states with cosmological
constant consistent with observation will have living organisms in them to
observe the value of the cosmological constant.  Finally, they argue that
their models solve the empty universe problem encountered in several
previous attempts to find an anthropic resolution of the cosmological
constant problem.

The anthropic determination of the constants of nature is very
controversial\footnote{Weinberg has remarked that a physicist
talking about the anthropic principle runs the same risk as a
cleric talking about pornography: no matter how much you say
you're against it, some people will think you're a little too
interested.}.  The present paper may be read as a critique of
anthropic arguments in general; it was directly inspired by the
work of Bousso and Polchinski, which is, in our view,
one of the more successful efforts to date to give
a framework for such an anthropic discussion.  We begin with a statement of
principle about the anthropic principle, and argue that within a
certain context one might imagine a scientifically defensible
anthropic determination of the cosmological constant, but that
with the current state of scientific knowledge, it does not make
sense to discuss anthropic arguments for other physical
parameters. Bousso and Polchinski proposed various scenarios to
obtain a small cosmological constant. All require utilizing the
strong version \cite{gvil} of Weinberg's galaxy formation bound.
In one, in which the compactification scale does not differ
significantly from the fundamental scale, we find that the gauge
hierarchy problem of the standard model is not resolved.
Since, within our groundrules,
one cannot explain this fine tuning
anthropically, the model fails to be a consistent theory of our
universe, but rather predicts that among those points in the BP
discretuum which might contain intelligent organisms, the typical
point does not contain a low energy spontaneously broken gauge
theory.

In scenarios with $d$ large compact dimensions, different issues
arise.  One cannot consistently ignore the flux dependence of the
potential for moduli.  As a consequence,
if $d \ne 2$, cancellation of the
cosmological constant still requires significant fine tuning.  For
$d=2$ \cite{arkani}, however, it appears possible, in principle, to cancel
the cosmological constant and to obtain a suitable hierarchy.  This
picture is much like that of the large dimension picture of
\cite{arkani}, where the logarithmic behavior of massless
propagators in two dimensions can give rise to hierarchies.

We have found another mechanism, which employs six-branes wrapped on
CY three-cycles, which does not require large internal dimensions
and achieves results similar to those of the two-dimensional BP
models, in what appears to be a more generic region of moduli
space.

We then describe a rational version of the irrational axion model
\cite{bds} which might be derivable from string theory.   This
model can solve both the hierarchy problem (using SUSY) and the
cosmological constant problem (anthropically).

The empty universe problem often arises in discussions of
solutions of the cosmological constant \cite{a} \cite{brt}.
We believe that this is an artifact of constructing one's theory
out of noninteracting modules.  In a realistic theory, the
potential for any inflaton field can have a dependence on the
discrete vacuum labels.  As a consequence, as long as the inflaton
potential does not have a large flux independent piece, after the final
tunneling event, the inflaton begins its classical motion
at a point far removed from its vacuum value.  It can thus reheat
the universe in the standard manner\footnote{The idea that the
excitation of an inflaton away from its vacuum value could be a
consequence of tunneling, was first employed in
\cite{tbone}-\cite{tbtwo}.}.
However, in models of inflation with only very small fine tuning
of parameters, this proposal leads, as a consequence of the low
scale of the inflaton potential, to too small an amplitude for
density perturbations.  Thus, all of the models which solve both
the gauge hierarchy and cosmological constant problems, either
have an empty universe problem, or predict too small an amplitude
for density perturbations, or require finely tuned models of
inflation.

We find this disturbing in the context of models based on a large discretuum,
because all of the low energy parameters are random variables.
Indeed, there are generically many solutions to the anthropic
constraint on the cosmological constant, all of which have very
different low energy physics.  Since we discount the possibility
of explaining the values of other parameters anthropically, we are
left in an uncomfortable situation if we find that generic models
in the class do not give correct predictions for low energy
physics.

We should mention a very recent proposal to solve the cosmological
constant problem using four-form fluxes which invokes, at most,
much more mild applications of the anthropic principle.
This idea involves tunneling through
a large set of metastable states, a tunneling which finally
ends when the cosmological constant is very small.
If successfully realized, such a proposal would not suffer
from many of the problems to be discussed here\cite{fmrsw}
(though some of the cosmological difficulties might persist).
We will not explore this proposal here.

Finally, we propose a rather different, and much
more modest, application of anthropic
arguments.  We imagine that generic SUSY violating vacua of
M-theory are unstable and/or do not lead to any sort of large
spacetime.  There might be only a few special points which are
dynamically stable once SUSY is violated.  Even if one of them
resembles our world, we must still understand why Nature did not
choose one of the stable SUSY vacuum states.  We suggest that
anthropic arguments may provide part of the answer to this
question, and give a broadbrush description of how such an argument might work.

As we were completing this paper, the paper \cite{jondon} appeared, which
has some overlap with our considerations.

\section{\bf A principled approach to the anthropic principle}

In our opinion, at the present
time, any scientifically defensible use of the anthropic
principle must conform to the following two guidelines:
\begin{itemize}
\item It must be embedded in an explicit mathematical model which truly
has an enormous number of ground states,
including some with the observed values of low energy coupling constants.
\item It must not make arbitrary assumptions about the necessary conditions
for intelligent life.

\end{itemize}

As to the first point, we will see that the Bousso-Polchinski analysis is the first
plausible demonstration that the required dense set of metastable
states might exist in string theory, but it is certainly
far from convincing.
As to the second, we will, for most of our discussion,
adopt the viewpoint that the anthropic principle requires
only that it be possible for complex forms of life
to develop.  We will only briefly touch on scenarios in which
it might make sense to restrict one's attention to carbon based life.
Then, given the current
state of physics and biology, we claim that the
second guideline rules out, for the forseeable future, the possibility of an
anthropic explanation of any parameter besides the cosmological
constant. The current state of biology is that we do not have an
explanation of our own carbon/oxygen based form of life.  The best
expert scientific estimates of the probability for finding another
intelligent species in our galaxy differ among themselves by many orders of
magnitude.  The current state of physics is that we do not have
any fundamental explanation for what the low energy gauge group
and its matter representations are.  A consistent anthropic
argument for {\it e.g.} the fine structure constant would have to
show not only that the value of the fine structure constant was
necessary to the existence of a large number of intelligent
civilizations composed of carbon/oxygen people, but also that
there were no other chemical compositions of life or different low
energy gauge theories whose life forms differed from our own by
virtue of having different low energy physics, which gave rise to comparable
or larger numbers of civilizations.
Thus, an anthropic argument which explains the value of some
parameter makes sense if vacuum states with different values of
this parameter suffer catastrophes so universal and cataclysmic
that we can be absolutely sure (on the basis of current physics)
that they can contain no intelligent life forms. The only
parameters for which this seems conceivable are the cosmological
constant, and (perhaps) the dimension of spacetime.   They affect
the universal dynamics of gravity and may thus lead to highly
generic phenomena.  We emphasize that within particular scenarios
in particular vacuum states one may find that other parameters can
affect the history of the universe in what appear to be
catastrophic ways.  For example, in states whose physics is close
to what we observe, various parameters may affect the baryon
asymmetry by many orders of magnitude.  However, choosing the
value of such a parameter anthropically would violate our
principles unless we could show conclusively that no other states of the
theory exist which have radically different low energy physics, but
a reasonable probability of having some form of life.\footnote{We 
emphasize that it is conceivable that with a more sophisticated
understanding of physics and biology than we possess today, we
might eventually find that the explanation of the value of some
of the constants of nature would depend on the details of
 {\it e.g.}, carbon based chemistry. However, for the forseeable
future, this is not likely to be a question we can address
scientifically.}

Our first
guideline forces us to work within the context of a real theory
with many ground states.  The only candidate we have for such a
theory at present is M-theory, and the candidate ground states in
this theory include some with low energy physics wildly different
from the Standard Model.  Once we have accepted the anthropic
principle we must, within M-theory, show not only that the most
probable ground states with life resembling our own must have a
certain parameter tuned to get the right baryon asymmetry, but
also that this vacuum does not have a much lower probability for producing life
forms than any other M-theory vacuum (or that it is dynamically
chosen by minimizing the potential).  Again, this is a task far
beyond our current abilities.  The only acceptable context for an
anthropic determination of the sort of parameter under discussion,
would be a theory with a large number of ground states which all
had standard model low energy physics and differed only in the
value of certain parameters.  Prior to the work of BP, we would
have found this possibility totally implausible. In the context of
the BP approach, the question of whether there are many vacua
with standard model low energy physics depends on how the moduli
are affected by the fluxes.

In a geometrical picture, nonabelian gauge groups arise from
singularities on moduli space.  If the moduli which control the
singularity have nothing to do with the cycles which bear the BP
fluxes then it is perhaps plausible that we could have the same
low energy gauge groups for many BP vacua.  The parameters in the
low energy gauge theory would be vacuum dependent\footnote{In the
large dimension scenario discussed below, it is possible that the low
energy parameters depend only weakly on the fluxes.}.  In this context,
we might imagine determining some other parameters anthropically.
For example, it is plausible (though certainly not proven) that,
within the standard model, the existence of life depends on the
existence of a baryon asymmetry.  Given a unique and verified
theory of how the asymmetry is generated (which we do not yet
have, and which probably depends on the existence of degrees of
freedom outside the standard model) one could then get an
anthropic constraint on the standard model parameters by insisting
that a baryon asymmetry of some reasonable size was generated.
This is likely to constrain the parameters to lie within a (not
terribly small) distance from a submanifold of low codimension.

Although we have now identified a set of hypotheses under which
anthropic determination of some other parameters would be valid,
the price to be paid is high: in such a model, none of the
parameters of low energy physics is determined by anything other
than these very weak anthropic constraints.  An
honest evaluation of such a model would require us to determine
the properties of all the vacuum states within the anthropic range
and see whether the values of parameters which fit our
observations are not highly improbable members of the
distribution.  If they are, one would have a fine tuning
problem.  There are many parameters in the standard model
(masses and mixings of fermions in the heavier generations) which appear
somewhat fine tuned but are unlikely to significantly affect
the question of whether there is a baryon asymmetry, so it seems likely
to us that even with the hypothesis that all BP vacua within the
anthropic range of the cosmological constant have standard model
physics with varying parameters one is unlikely to get
a satisfactory anthropic explanation of the world.  This is disturbing,
because we will find that there are typically many BP vacua which
satisfy the anthropic bound on $\Lambda$, so that this sort of model
forces one to find anthropic explanations for the other constants of nature.

There is a second approach to the sort of anthropic argument
described by BP, whose existence we feel we must acknowledge. One
can imagine a vast multiverse in which all of the metastable vacua
are realized at different ``places''. Then the much more
restrictive anthropic arguments based on our own biochemistry
could be used to explain why we were in a particular place. One
would need only to show the existence of vacuum states with low
energy parameters in the right range and would not have to take
into account questions of how typical these were among all vacuum
states which could support some kind of life.

It seems to us that this scenario amounts to abandoning the basic
goal of fundamental physics, which is to find the principles which
regulate the behavior of the world we observe.  That quest, in
the kind of theory envisaged in the previous paragraph, would stop
at the standard model.  All further explanation of the world we
see would have a contingent nature, like the explanations of
history and geography.  The fundamental theory of physics would
mostly be about things we can never in principle observe.  In
addition to our evident distaste for such a state of affairs, we
have serious doubts about whether it is compatible with the
holographic ideas which seem to play a central role in M-theory.
The different ``places in the multiverse'' are outside each other's
event horizons and we are ascribing independent degrees of freedom
to each of them.  Though we cannot rule it out at present, we
suspect that such a scenario is incompatible with the holographic
principle.

One other disturbing feature of such a picture should be noted.
In such a model, all of the constants of nature relevant to low energy physics
are random variables of this sort. As we have remarked above,
it seems highly unlikely
that they are all fixed by anthropic considerations.  Thus,
for example, one might imagine that the mass of the $u$ quark
or the value of the fine structure constant is fixed by anthropic
reasoning, but it seems unlikely that, say, all
of the values of neutrino
masses, or all of the elements of the KM matrix and the heavier
quark masses,
can be fixed in this way.  Thus, even the strong anthropic principle
implied by the multiverse picture will, in the end, leave us with fine
tuning puzzles that cannot be resolved.

To our knowledge, the first attempt to make an anthropic
determination of the cosmological constant incorporating our first
guideline was \cite{tbone}-\cite{tbtwo}.  This model combined a
scalar field called a {\it relaxon} with an extremely flat
potential (which could be justified using a variety of symmetries)
with a (then) conventional tunneling inflaton field. The inflaton
was supposed to tunnel out of  a false DeSitter vacuum and reheat
single bubbles.  As a consequence of the exponential expansion,
the bubbles never coalesced and each was a potential universe.
This process continued eternally.  The low energy cosmological
constant in each bubble was determined by the position of the
relaxon field at the time the bubble was nucleated. This gave rise
to an ensemble of universes with various values of the
cosmological constant.  It was conjectured that only those with a
cosmological constant within a few orders of magnitude of what was
then the observational upper bound would be able to support life.
This would have been an adequate explanation of the observations
if the distribution for $\Lambda$ had been flat.  But in fact,
according to the logic of the model, the number of possible
universes grows exponentially with time, as $\Lambda$ decreases
and so there is a prediction that $\Lambda $ take on the largest
possible negative value compatible with the existence of life.  It
was later argued \cite{wone} that this might actually be zero
within the observational errors. However, today the model is
surely ruled out.

Abbott \cite{a} attempted to rescue this model by adding a rapidly
oscillating potential for the scalar field, which had many minima
and in particular, one near the origin.  The universe was then
argued to sequentially tunnel, rather than slowly roll, down to
the last positive minimum.  This minimum is stable because of the
Coleman-DeLuccia suppression of tunneling into Anti DeSitter
spaces. The problem is that in the penultimate metastable state,
the universe inflates away all of its energy density and finds
itself in a state with very small positive cosmological constant,
but no matter or radiation.  This is the {\it empty universe
problem}.

Weinberg attempted to constrain models of this type by doing the
first honest calculation of the anthropic bound on the
cosmological constant.  He pointed out that once galaxies form,
there can be no further bound. Given a single galaxy one will
inevitably have stars and planets, and physics inside this
gravitationally bound system will be insensitive to  the
exponential expansion of the universe on time scales much longer
than the age of our universe.   Weinberg's original estimate was a
bound of $10^{2}-10^3$ times the critical density. Recent work has attempted
to argue that the {\it typical} value of the cosmological constant in an
ensemble satisfying Weinberg's bound is actually of order the critical
density \cite{gvil}.
The arguments which go into this
bound depend only on the physics of gravity and the equations of
state of matter and radiation. Thus, it has the insensitivity to
our ignorance required by our second guideline\footnote{Weinberg's
bound does implicitly assume that galaxies are necessary for life.
The eminent astronomer F. Hoyle once suggested
\cite{accordingtohoyle} (albeit in a work of speculative fiction)
that the typical lifeforms in the universe were vast interstellar
gas clouds who were very surprised to find that life could
exist on planets. It is clear to us neither that Weinberg's bound
would apply in such a hypothetical situation, nor that we know
enough about complex systems to rule out Hoyle's Conjecture.  Such
are the travails of anyone foolish enough to think seriously about
the anthropic principle.}.  In the next section we will see
whether the models of Bousso and Polchinski are similarly
insensitive.

\section{\bf Flux redux}

The work of BT was based on the observation that screening of 4-form 
flux by membrane antimembrane creation provided a mechanism
for dynamically changing the cosmological constant in a four
dimensional universe.  This is a precise analog of screening of
electric fields by charged particle creation in the Schwinger
model.  BP generalize this idea by observing that M-theory
compactifications on a 7-manifold with many 3-cycles (we will
generally think of a Calabi-Yau 3-fold fibered over a circle or an
interval), will have many four-form fluxes in the four
dimensional effective field theory. The corresponding membranes
are the original membrane of M-theory and five-branes wrapped
around the three-cycles. If there are $N$ such fluxes, the low
energy effective energy has an energy density, in the case
of a torus
\eqn{fluxen}{ {M^6
\over M_P^2} \sum_1^N q_i^2 n_i^2 - \Lambda_b} where $q_i^2$ (in a
slight change of notation from BP) is the volume of the i-th 3-cycle in
fundamental units ($V_i M^3$)
 and $\Lambda_b$ a bare cosmological constant independent of the fluxes.
 $M_P$ is the four-dimensional Planck mass, given by the equation
$M_P^2 = M^2 (M^7 V_7)$. In the simplest scenario.  While this
equation looks rather general, this is deceptive. It is valid for
the seven torus, but not for manifolds in which some cycles can
shrink to zero at fixed volume, with the volume of Poincar\'e dual
cycles also fixed.  BP argue that if the $q_i$ are
incommensurable, and if either the unit of energy density involved
in the change of a single flux is small compared to $\Lambda_b$ or
the number of fluxes is large, then there will always be points in
the {\it discretuum} of allowed values of $n_i$, for which the
cosmological constant is consistent with observation.  These vacua
are then chosen by the anthropic principle.

There are a large number of assumptions involved in writing the
energy equation in the simple form above (most of these are
mentioned by BP). First and foremost, we are using low energy
effective field theory.  Thus, all energy densities should be much
smaller than the fundamental scale, and fluxes should not be so
large that higher order terms in the effective action are
important.  As an example of what is involved, note that in 11D
SUGRA, the BP fluxes give tree level masses to many components of
the gravitino.  The one loop Coleman-Weinberg formula for the
dependence of the vacuum energy density on the fluxes contains
terms quartic in the flux.  Thus to trust the BP analysis one must
be in a regime where these and other terms are negligible. BP
achieve this by insisting that some of the internal dimensions are
larger than the fundamental scale.  We can distinguish
two cases:  the dimensions are slightly larger than the fundamental
scale, or they are much larger.

Consider, first, the case that they are slightly
larger. Note that low energy
supersymmetry, in the conventional sense, is not likely to be
relevant in this case,
since the energy associated with the fluxes will be far larger
than the would-be scale of supersymmetry breaking, and the fluxes
themselves break SUSY.  If the dimensions
are only slightly
larger (up to a few orders of magnitude), then the parameters of the
standard model, such as the Higgs mass and the value of the QCD
scale, as well as gauge and Yukawa couplings, will be sensitive
to the values of the fluxes.  Moreover, there is likely
to be a huge number of states which satisfy the anthropic
bound on the cosmological constant.  For example, if there
are, as in the discussion of BP, of order $120$ fluxes,
than changing the values of the charges ($q_i$) by
a factor of $2$ changes the number of acceptable
states by a factor of order $10^{60}$.  So all of the parameters of the
standard model---and in particular
the gauge hierarchy---will be, unpredictable,
simply determined by anthropic considerations.
The Higgs mass, for example, will receive contributions from
couplings to the four-form flux, such as
\begin{equation}
\int d^d y d^4 x \sqrt{g} F_{MNOP}^2 \vert \phi \vert^2
\end{equation}
Corrections to the gauge couplings, from
operators such as $F_{MNOP}^2 F_{\mu
\nu}^2$, will be nominally of order one ($\int d^dy \sqrt{g} F_{MNOP}^2$
is, by assumption, of order the fundamental scale); similar
remarks apply to the Yukawa couplings.

This violates the principles we laid out earlier.  Note also
that one is also making the assumption, here, that any moduli of the
compactification (e.g. those associated with the relatively
large size of the compact dimensions) have stable minima for
all values of the fluxes.  This seems a quite strong assumption,
given that the flux contributions to the potential are not likely
to be much smaller than other contributions.

We have spoken of the Higgs particle, in this small extra dimension
case, but one could also imagine that technicolor plays some
role.  However, in the anthropic context, technicolor scenarios are highly
suspect, if not ruled out altogether.  Even the most ardent
enthusiast would admit that the generic technicolor model is
inconsistent with experiment.  To promote an anthropic technicolor
scenario consistent with our rules, advocates would have to show
that these models not only arose among those BP vacua with
reasonable values of the cosmological constant, but also that
technicolor models consistent with experiment were generic in
these vacua.  Otherwise, and this is the most likely conclusion,
the BP prediction for a universe with relatively small internal
dimensions is that the universe does not contain a spontaneously
broken gauge symmetry at the electroweak scale, or at best that it
does contain one but that the model is inconsistent with
observation.
%


The alternative,
is to consider the possibility that some dimensions are very
large, as in the large dimension proposal of \cite{arkani}.  In this case,
the assumption that the values of the moduli are independent of
the fluxes is almost certainly not correct, and
it is most natural to imagine that the fluxes play a role in
fixing the moduli.
We will first
assume that there are $d\geq 3$ dimensions with $RM \gg 1$ and,
for the moment, that the cycles which define the fluxes have
linear scale of order $R$. Most other assumptions about the number
of large dimensions and the flux bearing cycles lead to similar
conclusions. We will explore some possible exceptional cases
individually below.  We further assume that the standard model
lives on a brane, that the bulk is supersymmetric and that SUSY breaking
comes from a brane, not necessarily the standard model brane.  In
particular, the leading order in $RM$ bulk geometry will be
assumed Ricci-flat.

Contributions to the four-dimensional cosmological constant can
now be classified by their dependence on $RM$.  A brane
cosmological constant is of order one.  A bulk curvature squared
term is of order $(RM)^{(d-4)}$ and there are various bulk
contributions to the energy which scale like $(RM)^{(d-6)}$. The
flux terms scale like $(RM)^{(d-6)}$.  They are thus subdominant
at large $RM$ but we can make them arbitrarily large and positive
by increasing the value of the quantized flux (though we should
stop before violating the rules of effective field theory).
Finally, we can imagine some independent flux which (perhaps
unrealistically) we assume cannot be changed by dynamical
tunneling processes. This will give a positive term in the energy,
with a coefficient that we can allow to be large. We will call it
the {\it flux ex machina}.

Assume first that $d > 4$.  Then the energy is dominated by the
curvature squared term, $a (RM)^{(d-4)}$.  In order for the
potential to be bounded below at large $RM$, $a$ must be
positive\footnote{We note that a metastable minimum for $RM$ would
be incompatible with the BP idea.  In the BP scenario the universe
is imagined to tunnel around the flux lattice many times before
finding the anthropic vacuum with small cosmological constant.  No
metastable vacuum for the moduli could have a long enough lifetime
to ``wait for the BP process to be completed''.}.   The radius can
be stabilized at a large value by balancing the contribution of a
large BP flux against this curvature squared term.  However, both
terms give positive contributions to the cosmological constant and
the BP cancellation is not operative.  The situation is not
improved by introducing the {\it flux ex machina} since it too
gives a positive contribution to the cosmological constant.  Note
that it does not make sense to stabilize the radius by balancing
the positive curvature squared term against, {\it e.g.} a
curvature cubed term.  These can balance at large $RM$ only if we
fine tune dimensionless coefficients.  One would have to find a
theory which had a variety of vacuum states for each value of the
BP flux, in which the short distance physics induced many
different values for the coefficients in the effective Lagrangian,
including finely tuned ones, in order to find an anthropic
explanation of this tuning.

We thus assume $d \leq 4$, in which case the brane cosmological
constant dominates the energy density in the absence of large
fluxes.  This can be assumed negative.  The BP flux term must be
of the same order of magnitude as this term at the anthropic value
of the BP fluxes.  Since this term depends on $R$, it must be one
of the important terms in the radial stabilization equation. For
the same reason, it is also much larger than the curvature cubed
term (which scales the same way as a function of $RM$).  For four
large dimensions this means that we must rely on a {\it flux ex
machina} whose contribution to the energy grows like $(RM)^p$ with
$p> 0$ in order to stabilize the radius.  But there are no such
fluxes in four dimensions, so we conclude that while the BP
mechanism can work, the radius cannot be stabilized at a large
value and the model is inconsistent.   If there are three large
dimensions, we can stabilize the radius by balancing the curvature
squared term against the BP flux\footnote{ In order for this mechanism
to work, the brane cosmological term must be of order $1/R$.  This can be
explained by SUSY if the SUSY breaking scale is no bigger than the
flux terms.}.  The required coefficient of the
curvature squared term is however negative (since both terms
decrease with $RM$) and we are again stuck with a metastable
minimum for the radius.


We now relax our assumptions and turn to the case of two large
dimensions.  The seven-form flux of M-theory is integrated over
one small and two large dimensions in order to get the BP fluxes
in the noncompact dimensions.  The energy of a BP flux now scales
like $(RM)^{(-2)}$.  The dominant term in the energy is no longer
the brane cosmological constant, but a term scaling like $a\,{\rm
ln} (RM) $ coming from the infrared behavior of two-dimensional
massless propagators.  We will however include the brane
cosmological constant as well because logarithms are not that
large even when $RM \sim 10^{15}$.   Stabilization now requires
$\sum n_i^2 \sim a(RM)^2$ and that the coefficient of the logarithm
is positive.  The requirement that the cosmological constant
cancel is now ${\rm ln} (\sum n_i^2 ) \sim \Lambda_b$.  The bare
cosmological constant must of course be negative.

In order to solve the gauge hierarchy problem, we follow
\cite{arkani} and ask that $M \sim 1$ TeV, which,\footnote{Of course the
phenomenological constraints on models with two large dimensions
tell us that $M$ must be greater than about $50$ TeV.  One must
presume that this is still consistent with solving the hierarchy
problem.}
given $M_P \sim 2 \times 10^{18}$ GeV, means $RM
\sim 10^{15} $. This requires $\Lambda_b$ to be an order of
magnitude or two higher than we might have expected it to be but
still much smaller that one would have obtained in a similar
theory with no bulk SUSY. We can also estimate the contributions
to the vacuum energy due to loops of particles which obtain SUSY
violating masses from the BP fluxes.  These are smaller than or
equal to the flux energies themselves, so the model appears self
consistent.

There are several attractive features of this model.  The BP
mechanism gives a rationale for the stabilization by large flux
postulated in \cite{savasetal}.  Using anthropic logic, the large
flux vacua are the only ones we are likely to see. The two ideas
interact synergistically, for the large flux also facilitates the
cancellation of the cosmological constant.
In this picture, the parameters of the SM
are not very sensitive to the values of the fluxes, since
the fluxes are of order $1/\sqrt{V}$.  Finally, the problem
of the small radius picture, that there are likely to be vast numbers
of states with suitable values of the cosmological constant,
is significantly ameliorated here.   If the volume is large,
only a small number of fluxes are required in order to implement
the BP cancellation of the cosmological term.  So rather than,
say, $10^{60}$ states, one might easily imagine that there
are only $100$'s or $1000$'s of states compatible with the bound.

\section{Brane-antibrane separation}

We would like to present another scenario realizing the basic philosophy
proposed by BP. We will also generate a large number of metastable states
among which we find some vacua with a small value of the cosmological
constant, but the way we obtain the large number of metastable states
will be different.

Let us imagine a compact space (e.g. a Calabi-Yau space) with a $p$-cycle
of a small volume (imagine $S^3$ which shrinks to zero at the conifold
point of the moduli space). Let us furthermore assume that there are two
different places where this $p$-cycle shrinks to a small size: we
therefore deal with submanifolds $X,Y$ which are topologically equivalent.
Now we can wrap $N$ $(p+3)$-branes on $X$ and $N$ $(p+3)$-antibranes on
$Y$. Those branes and antibranes are extended in the large dimensions, all
the transverse dimensions are compact and therefore the total brane charge
must vanish. There can be several pairs of cycles
$X_i,Y_i$, $i=1,\dots n_{cycles}$ in the same way as BP use many different
types of fluxes. The number $N_i$ of the branes wrapped on $X_i$ plays the
role of the flux in the BP picture.
An easy toy model for visualizing this configuration is a torus
$T^2$ whose $a$ cycle shrinks to zero at two opposite points along the $b$
cycle. We see that
although conservation of the charge does not prevent the branes from
annihilating,
there is an energy barrier that makes such an event unlikely.
The brane (or antibrane) must move through the ``thick'' region where its
tension is necessarily large.

Otherwise the physics is similar to that of the picture
of BP. There
is a negative ``bare'' cosmological constant which is compensated by a
large number of branes and antibranes wrapped on the shrunk cycles $X_i$,
$Y_i$, each of which contributes a positive amount to the vacuum energy.
Sometimes---after a cosmologically long period of time---a brane
tunnels through the ``thick'' region of the Calabi-Yau space and
annihilates with its antipartner, reducing the number $N_i$ by one.
The relevant instanton consists of a bubble in three space in
which the compact dimensions of the branes and antibranes move together and
annihilate.
If there are enough types of six-brane, and/or if the energy
densities associated with them are sufficiently small, it is
possible to cancel off the negative cosmological constant to the
desired degree of accuracy.

There are several reasons why we consider this picture to be more
flexible than the picture of BP. First of all, the tensions of the wrapped
branes can contribute small amounts to the 4-dimensional effective vacuum energy
The natural small number can be
obtained from the size of the shrunk cycle. Furthermore in the previous
sections we showed that it is hard to ensure both stabilization of the
moduli {\it and} the cancellation of the vacuum energy in the BP
picture.

In the case of the wrapped branes, the contribution to the vacuum energy
from the wrapped branes depends mostly on the geometry near $X_i,Y_i$ and
not much on the overall size of the manifold. Therefore we effectively
decouple the problems of the stabilization and the vacuum energy:  the
branes on $X_i,Y_i$ are responsible for the
cancellation of the cosmological constant, while a different dynamics
solves the stabilization of the overall size of the manifold.

As far as stabilizing the moduli of the shrinking cycle is
concerned, it appears sufficient to put a three-form flux onto the
shrinking cycles on which the branes are wrapped.  The size of the
cycle is then determined by the ratio of this flux and the 6-brane
charge, and is naturally small if the ratio is small.  Of course,
this discussion is too glib.  We are trying to discuss
nonsupersymmetric compactifications of 11D SUGRA on a complicated
manifold with fluxes and branes, and the real dynamics is highly
nontrivial.  Our discussion here is merely suggestive of the
existence of a stable solution.  One point in favor of this view is that by
making the manifold large, we can control the scale of SUSY
breaking and study configurations that are close to supersymmetric
ones.

This might also have phenomological implications.  If the manifold
is large, and the standard model lives in the bulk, far from the
small cycles, SUSY breaking in the standard model will be
suppressed.   Thus, this kind of model can support hidden sector
SUSY breaking, perhaps with the conventional intermediate scale.

\section{Rationalizing the Irrational Axion}

Some time ago, Seiberg and two of the present authors described
another class of models with a discretuum of vacua \cite{bds}.
Although the main thrust of that paper was an explanation of the
strong CP problem, it was noted that the same model also might
provide an anthropic resolution of the cosmological constant
problem.  We would now like to present an updated version of that
model which is at least superficially compatible with M-theory.


It is well known that the F-theory region of M-theory moduli
space \cite{vafafetal,vafamorone,vafamortwo}
can give rise to vacua with a direct product
gauge group with large numbers of relatively large factors. Let us
assume such a compactification, with $4$ conserved supercharges
(up to low energy gauge theory effects, which will break SUSY) and
a product $G\otimes G_1 \ldots \otimes G_N$.  $G$ will be
asymptotically free and have SUSY breaking dynamics at a scale
$M_{SUSY}$.  The $G_i$ are taken to be $SU(N_i )$ groups with a
sufficient number of matter fields to make them infrared stable at
the fundamental scale.  However, we imagine that their $\beta$
functions are relatively small because {\it e.g.} the number of
flavors of fundamental chiral fields is near the critical value.
Furthermore, we imagine that there are low energy couplings
between the fields in $G$ and those in the $G_i$.  That is, we
have low energy SUSY breaking in the $G_i$ rather than gravity
mediated effects.  The SUSY breaking removes some $G_i$ matter
fields, making the infrared running of the $G_i$ couplings
unstable. Since the $N_i$ are large, all of the $G_i$ gauge
couplings will become strong at scales not too far below
$M_{SUSY}$.  We will approximate this situation by saying that all
the scales are approximately the same, and call the single scale
$m$.


Finally, we imagine that all but one of the moduli of the theory
have been frozen, either by a high energy SUSY and R preserving
superpotential, or by the effects of SUSY breaking.  The single
axion field is a periodic field which originates as a component of
some higher dimensional $p$-form gauge field.  If $a$ is the
canonically normalized four-dimensional axion, then $a/f_a$ is a
periodic variable with period $2\pi$, and this defines the axion
decay constant $f_a$.  $a$ has couplings to the gauge fields of
the form $(a/f_a )\sum Q_i$, where $Q_i$ the topological charge
density of the gauge group $G_i$.  Instanton effects in the gauge
groups will give rise to an axion potential with period $2\pi$.
However, if the process of SUSY breaking leaves some unbroken
chiral symmetries, and a corresponding set of massless nonsinglet
fermions, then spontaneous breaking of these symmetries will give
us a discretuum of vacua.  For simplicity, suppose that all
fermions besides the gauginos are lifted by SUSY breaking. Then
each $SU(N_i)$ gauge theory will have $N_i$ vacua due to gaugino
condensation.

The potential generated for the axion by $G_i$ now has periodicity
$2\pi /N_i $, so we obtain a formula for the total potential
\eqn{axpot}{V = \sum_i e_i M^4 P_i (a/f_a N_i )} where the $P_i$
have period $2\pi$ and the $e_i$ are numbers of order one. If the
$N_i$ are relatively prime, this function will have $\prod N_i$
minima.  If $\prod N_i \sim (M^4 / \Lambda)$ then there will
typically be at least one minimum with $V \sim \Lambda$. Thus,
again we have a discretuum with at least one vacuum which satisfies
the observational constraints on the cosmological constant.

We have neglected the contribution to the cosmological constant
from the SUSY breaking dynamics itself.  We can take this into
account by including the coupling of the axion to $G$.  If the $G$
theory has only a few vacua it will generate a potential of the
form $M_{SUSY}^4 P_G (a/f_a) $ where $P_G$ has a short period (a
few times $2\pi$).  Thus, it will have many zeroes within the
period of the potential (\ref{axpot}). In regions of size $\Delta
a/f_a \sim (M/M_{SUSY})^4$ around these zeroes the problem reduces
to the one we solved previously, so if $\prod N_i \sim (M_{SUSY}^4
/ \Lambda )$ there will be a minimum with a cosmological constant
of order $\Lambda$.


\section{The empty universe problem and the cosmic microwave$\!$
background (CMB)}
%

Anthropic models of the cosmological constant often suffer from
the empty universe problem \cite{a} \cite{brt}.  That is, while
they explain why there is a vacuum state with small cosmological
constant, their mode of accessing that vacuum leaves them with no
mechanism for generating the entropy of the cosmos.  In broad
terms, the problem is that if one achieves the anthropic vacuum by
tunneling from a metastable state with positive cosmological
constant, one appears to tunnel into an empty vacuum.  The model of
\cite{tbone}-\cite{tbtwo} shows that this is not an inevitable consequence
of
anthropic ideas.

BP present three different suggestions for solving the empty
universe problem.  The first exploits the Hawking temperature of
the penultimate metastable DeSitter vacuum to ensure that an
inflaton tunnels to a nonvacuum value.  This mechanism is unlikely
to work in a model in which the gaps in the discretuum are small
compared to the Planck scale.  We have seen that the latter
condition is necessary to a solution of the hierarchy problem. The
second suggestion uses properties of a particular class of
inflaton potentials.  The third (which BP attribute to suggestions
of Susskind and Thomas) is, we believe, much more generic.  It is
simply to recognize that in a model containing a discretuum, $\{
n\}$ and an inflaton field $\phi$, the effective potential $V(\phi
, \{ n\} )$ is not apt to be a simple sum of a discretuum
potential and an inflaton potential.  That is, the inflaton
potential will depend on which point in the discretuum one is
working at.  This is essentially the same point that we have made
about the $\{ n \}$ dependence of the potential for moduli.

It is important that the flatness of the inflaton potential be
generic and independent of the discretuum.  This is easily
achieved if the inflaton is itself a modulus \cite{bgberkooz}\footnote{Modular 
inflation models still require two or three
orders of magnitude of fine tuning, which is much less than most
alternatives. Some ideas for explaining this tuning can be found
in \cite{ds}.}.  In such a model the tunneling process from the
penultimate point in the discretuum to the anthropic vacuum will
place the inflaton at some point on the equipotential surface
whose energy (including derivative contributions) is equal to that
of the false vacuum.  The particular point on this surface is
determined by minimizing the action of the relevant instanton.
This point is in the basin of attraction of the anthropic vacuum
but definitely not at it, and in models where the potential is a
generic function of discretuum and inflaton there is no reason why
we cannot have inflation and reheating.

The real problem is in the details.  Models which solve the
hierarchy problem have a low value for the gap in the discretuum.
In the axion and wrapped brane models we can tolerate a vacuum
energy scale of $10^{10.5}$ GeV if SUSY breaking is communicated
to the standard model via gravitational effects. In the BP model
with two large dimensions and a fundamental scale of a TeV, the
vacuum energy scale is $1$ TeV.  Given a modular inflation model,
the generic prediction for the
amplitude of CMB fluctuations in such models will be too small to
be compatible with observations.

There is a sense in which this problem is more severe than the
empty universe problem we claim to have averted.  As far as we can
see, models of the type we are discussing could easily produce
situations in which we would be unable to rule out the possibility
of intelligent life.  That is, for a broad range of choices of
both discrete (gauge groups) and continuous parameters in our
models, we would find a matter dominated era of the universe in
which galaxies could form and there was a rich and complex low
energy physics.  Nonetheless, the generic prediction would be that
the amplitude of primordial CMB fluctuations was incompatible with
observations in our universe.  We would conclude that within our
class of models, a universe that resembled our own was very
improbable, among all those which might have life in them.

We would like to acknowledge that many inflation theorists would
consider our argument to be a defect of modular inflation models
but not of the general anthropic ideas we are exploring here.
There are inflationary models with a low vacuum energy scale and
adequate amplitudes for density fluctuations.  Few of these models
satisfy the usual field theoretic criterion of naturalness. They
contain small dimensionless parameters whose size is not explained
by any symmetries.   As in our discussion of technicolor above, we
consider this a more serious defect in the anthropic context than
otherwise.  If one is writing down a phenomenological
approximation to a unique theory of the universe, one may hope
that violations of naturalness will eventually be explained by
explicit calculations in the theory of everything.  In the
anthropic context, one must view all parameters as being drawn
from an ensemble consistent with the very weak anthropic
constraint.  Violations of naturalness would seem to occur with
very low probability.

\section{Anthropic vacuum selection in M-theory}

So far, we have considered a variety of anthropic proposals for
understanding the cosmological constant, and possibly other
parameters in the low energy effective action. These suggestions
have in common the idea that there are a vast array of stable or
metastable ground states of string theory with rather generic
properties. We have found all of these ideas troubling in some
respect.  Not least of all, it is not clear that such a discretuum
exists in string theory.  In the case where the charges are large,
one would require, for example, that there be a stable minimum of
the moduli potential for every value of the fluxes.  Yet changes
in the fluxes are not extremely small perturbations in the theory,
and it is not clear whether such a set of stable solutions would
exist.  Indeed, our experience in M-theory is that it is very hard
to stabilize the moduli once SUSY is broken.
In the case of large dimensions, we saw that only for
the case of two large compact dimensions is it possible to cancel
the cosmological constant without fine tuning.  Whether or not
there exists a discretuum of solutions depends on detailed
properties of the theory, which we are not in a position to
explore at the present time.

In this section, we propose an alternative viewpoint.  Rather than
arguing that there are innumerable string vacua with properties
almost compatible with the existence of life, we would like to
consider what we view as a far more reasonable possibility:
generic string vacua are incompatible with the formation of
structure in the universe.  This, we would suggest, might be the
solution of the vacuum degeneracy problem of string theory---provided 
that there are at least a small number of vacua
compatible with the very weak anthropic principle we have advanced
in section 2.

To understand this point, let us divide the known ground states of
string theory into categories:
\begin{itemize}
\item  Supersymmetric vacua in dimension $d \ge 5$, or
 supersymmetric vacua in dimension $d =4$ with more than eight
supersymmetries.
\item  Supersymmetric vacua in dimension $d=4$ with four or
eight supersymmetries and exact moduli (as well as some
approximate moduli).
\item  Four-dimensional vacua with SUSY restored only in extreme regions
of moduli space
where gravity decouples.
\item  Vacua in $d=4$ without supersymmetry anywhere in the
moduli space---{\it i.e.} with Planck scale SUSY breaking.  Note
that most of the hypothetical BP vacua fall into this category.
\item  Theories in $d < 4$ with or without supersymmetry.
\end{itemize}

Our goal will not be to prove that any of these
possibilities is incompatible with the development of a large
universe with structure on interesting scales, but rather to argue
that it is plausible that generic vacua in each of these classes
are incompatible with the very weak version of the anthropic
principle we have proposed.  This is, in our view, the most
reasonable possibility, compatible with our present understanding
of M-theory, for understanding the problem of vacuum selection.
The reasons our arguments must be very tentative are easy to
understand.  First, if we give up the anthropic explanation of the
cosmological constant, we don't know how (or whether) the
cosmological constant problem is solved in string theory.
We might imagine a gamut of possibilities: that it is only solved in states with
some degree of (approximate) supersymmetry, that it is always solved, or that it
is never solved.  Second, we have no understanding of the degree
to which inflation is generic in string theory.  Our working
assumption is that it is reasonably generic, i.e. that it
happens in some finite (if small) fraction of string
ground states.  For example, if the scale of some modulus
potential is lower than the string scale, and if the modulus
varies on a scale of order $M_p$, then the number of $e$-foldings is
of order one; presumably if it is of order $20-30$, one has the
possibility of structure formation.

With the understanding that our goal is only to provide suggestive,
rather than definitive, arguments, let us consider the various
cases in turn.
\begin{itemize}
\item
Theories in $d > 4$, with some degree of supersymmetry, or in 4
dimensions
with more than 8 supercharges:  In all of
these theories, there are exact moduli spaces.  The dynamics on
the moduli space is highly constrained.  Suppose that the universe
was at some time hot, and reasonably large.  Subsequently, the
energy density in the zero modes of the moduli redshifts much more
rapidly than $T^d$.  Furthermore, there is an instability
\cite{bgberkooz}
which converts modular zero mode energy rapidly into radiation.
There are
stable massive objects in these models, but they are BPS and in generic
regions of M-theory moduli space have mass of order the Planck scale.
The only conserved quantum numbers they carry are coupled to long range
fields and there cannot be an excess of particles over antiparticles.
There are no known metastable neutral particles.

There is no mechanism for producing these particles in the regime
where the temperature is well below the Planck mass and
semiclassical cosmology makes sense.  Thus, their initial density
is an input. However, we can conclude that structure formation is
unlikely for any initial density.  Would be structures are
composed of equal mixtures of particles and antiparticles and
rapidly annihilate into radiation. It seems that generic models of this type are radiation
dominated throughout all of cosmic history, and no structures are
formed.

We realize that there are loopholes in these
arguments, but suggest that it is likely that they can be closed as
our understanding of these states improves.

\item
Theories in $d=4$ with approximate $N=1$ supersymmetry in extreme
regions of the moduli space:   It is well known that these
theories tend to exhibit runaway behavior towards supersymmetric
regions of moduli space. No example is known where one can show,
reliably, that there is a stable local minimum of the potential in
the interior of the moduli space.  While in general we do
not expect metastable minima in the interior of the moduli
space, it is plausible that there
are reasonably stable minima in some cases.  The question would be
how frequently such minima appear. Roughly speaking, we might imagine local minima are
metastable when, for example, there is a low energy theory
with moderately small gauge couplings.   We do not know how to
assess this possibility, since we have no examples.
It is possible that they are rare.  Racetrack models have been
proposed as a mechanism to stabilize some moduli \cite{ds}.  These models
involve discrete fine tunings of gauge groups and particle
content, and so, by definition, if they occur at all
they are rare.  We will assume that this is
the case; generic string states with asymptotic $N=1$ SUSY exhibit
only cosmological solutions.  Can these be compatible with the
formation of structure?  Quite generally, for the superpotentials
which give rise to runaway behavior, the potential is related to
the scale of supersymmetry breaking.  Generally for these
potentials the potential energy drops more rapidly than $1/R^4$, and
the temperature always dominates the energy \cite{bgberkooz}.

\item
Theories where SUSY is broken at the Planck scale and there is no
runaway to a supersymmetric region.  All examples of this type
have negative potentials which draw the system towards apparently
catastrophic regions in the interior of moduli space.  Since
reliable approximations break down before the catastrophe is
achieved, we cannot say for certain that these models have no
sensible large scale physics, but it appears highly unlikely.
Note that once one puts moduli into the picture, the BP vacua are
in this category.  In the bulk of the text we have imagined, with
BP, the possibility that most of the moduli are stabilized by
some higher energy dynamics.  Here we opine that this is unlikely,
and that the BP vacua are not really stable.

\item
Theories in $d<4$:  Here the bizarre nature of long range
gravitational interactions makes it unlikely that structure will
form.  There is no Newtonian gravitational potential, no Jeans
instability, and no reason to expect galaxies.

\item
The most likely candidate vacua in which there might be structure
formation but obvious disagreement with observation, are those
with exact $N=1,2$ supersymmetric moduli spaces in $d=4$. ($N=1$
theories with exact moduli spaces were discussed in
\cite{bdexact}.)  These
vacua can have weakly coupled gauge theories in them which
generate a variety of scales well below the Planck scale. As a
consequence, it is possible to have approximately conserved
quantum numbers like baryon number or exact
discrete symmetries which guarantee the existence
of long lived massive particles (this
was not the case in the examples described in
\cite{bdexact}, however). Given this possibility, depending
on initial conditions and the details of the dynamics, one may
develop an asymmetry in the approximate quantum number. Stable
structures can now develop via gravitational clumping. Note that
the cosmological constant vanishes in such scenarios, so something
like galaxy formation may very well occur.  For the $N=2$ case, the
region of moduli space where there is strong IR dynamics at scales
well below the Planck scale is very tiny, and one may argue that
these vacua require fine tuning.  However, for $N=1$ there may be
codimension zero moduli spaces with unbroken nonabelian groups.
The real question with regard to these vacua is how likely it is
that one has weakly coupled gauge theories.  This is a question
that we do not yet know how to answer.  One possibility for $N=1$
is that the gauge couplings depend on the exact moduli. If the
universe is closed, the exact moduli will be time dependent early
in cosmological history, but will come to a halt at some point in
moduli space.  The question of whether gauge couplings are weak
would depend on the initial conditions.  Thus, the search for an
anthropic vacuum selection principle might fail because of our
inability to rule out vacua in this category.  While it is
possible that SUSY is incompatible with life, it seems possible
that it is compatible with structure formation.

\end{itemize}

Our proposed anthropic vacuum selection principle can rule out
large classes of string vacua, but vacua with exact $N=1$ SUSY
resemble the real world a little too closely to be eliminated by
the very weak form of the anthropic argument that we have
advocated.  We have however assumed that reasonably weak gauge
couplings are easy to come by in M-theory.  Perhaps this is not
the case.  One way of explaining the weakness of gauge couplings
utilizes the racetrack mechanism \cite{ds}.  This requires
an intricate pattern of gauge groups and might be realized only
rarely among solutions of M-theory.  Perhaps, for reasons we do
not currently understand, it cannot occur in vacua that are also
supersymmetric.  Or perhaps exact $N=1$ moduli spaces with
metastable particles are, for  some reason, very rare.
Otherwise, one would have to hope for a more
detailed anthropic argument that could eliminate exact SUSY.  At
the moment, we do not have one.

\section{Conclusions}

We have examined anthropic models which explain the value of  the
cosmological constant in terms of tunneling between a large
discrete set of metastable vacua. Within the modern M-theoretic
point of view, we emphasized the necessity of considering modular
stabilization in conjunction with minimization of the vacuum
energy with respect to discrete parameters.  Examining the bare
minimum in this regard, namely stabilization of the overall
breathing mode of the internal manifold, we found strong
constraints on such models.  In particular, when we incorporate
the further requirement of solving the gauge hierarchy problem, we
find that the only sensible models have a vacuum energy scale
bounded by about $10^{11}$ GeV.  Such models generally lead to a
prediction for the amplitude of CMB fluctuations which is orders
of magnitude too low.

We presented some ground rules for anthropic models which
incorporate a decent respect for our ignorance of the physical
basis of intelligent life.  These rules imply that in our current
state of knowledge, anthropic determination of parameters is
acceptable only if the physics involved in the argument is generic
and model independent, and if the disaster implied by values of
the parameters outside their experimentally determined range was
sufficiently terrible for us to conclude that no form of 
self-organized behavior would be possible.  In our view this argument
virtually rules out anthropic determination of parameters other
than the cosmological constant and (perhaps) the number of large
spacetime dimensions.  For example, we believe that strong anthropic
determination of the vacuum state in M-theory (we must have an
$SU(3,2,1)$ gauge theory in order to have conventional chemistry
and carbon based life, {\it etc.}) would not be scientifically
defensible.

We would also like to
reiterate a remark about the anthropic models studied in this
paper which is perhaps the most disturbing criticism of this circle of
ideas.  Our initial attitude towards anthropic arguments was
that they might be reasonable when restricted to the cosmological
constant.  However, the central point of models of the BP type is
that we can explain an extraordinarily small number in terms of a
concatenation of numbers of reasonable magnitude.

As we have seen, if the internal dimensions in the BP scenario are
small, or if one adopts the almost irrational axion models,
one finds that there will be {\it many} metastable
vacua with cosmological constants within anthropic bounds.
Insisting that there be only one would be an artificial and
arbitrary constraint on the model, with not even anthropic
justification. The set of vacua within the anthropic window for
the cosmological constant will differ from each other in many
ways. Thus, all of the details of low energy physics (values of
constants in the low energy Lagrangian and perhaps even its field
content) must be viewed as being chosen from an ensemble. These
models thus require anthropic arguments to explain not just the
cosmological constant, but also the other parameters of the low
energy effective Lagrangian. If we are to declare such a model an
acceptable explanation of the physical world, we must decide
between the two philosophies we have sketched above.

According to the first we would have to show that the vacuum in
which the details fit experiment was typical among all those for
which we could not rule out the possibility of life.  Thus, the
work involved in verifying such a model of physics is multiplied
many times compared to our usual task of simply calculating all
the numerical details of a complicated dynamical system and
comparing them with experiment. We must repeat the calculation for
all the other vacua (with no guide from experiment) and show that
our effective Lagrangian was typical.  We are more likely to be
able to show that it is not, and to rule the model out.  And even
if we succeeded in this herculean task we would be left with the
conclusion that the detailed numerical values of low energy
parameters were a statistical accident with no hope of scientific
explanation. It is a very bleak picture of the future of physics.

The second philosophy
declares that all vacua compatible with the weak anthropic
principle are actually realized, in a vast multiverse whose
component universes can never communicate with each other. We are
just one of many forms of life in the multiverse, and we can use
the strong anthropic principle to explain why our particular life
form lives in the component of the multiverse where the parameters
are {\it just so}.  The fundamental theory has no more to say
about all the low energy physics we will ever observe than that it
is a possible metastable state of the theory (very likely a very
special one). We leave further discussion of this philosophy to
its proponents.

We have seen that in the case of two large dimensions, the situation
is better.  The parameters of the SM are not so sensitive to
the values of the fluxes, and, while the number of vacua
compatible with formation of structure is probably large, it
need not be enormous.

Thus,  we believe that it may be possible to overcome the
difficulties of the anthropic models that have been discussed so
far.
Still, we are extremely skeptical of this mode of explanation. This
skepticism is scientifically based, but coincides (perhaps
suspiciously) with our emotional reaction to the notion of an
anthropic resolution of the cosmological constant problem. Puzzles
like that of the magnitude of the cosmological constant often lead
to revolutionary conceptual changes in the structure of physics.
There is a wealth of vague and partial evidence that our notion of
how M-theory reduces to local field theory is flawed. This
evidence relates to the holographic principle and the idea that M-theory
has many fewer degrees of freedom than field theory
(exactly the opposite of the naive counting).  It is obvious that
these ideas have a bearing on the cosmological constant
problem and it seems unlikely to us that the problem will be
solved by anthropic reasoning in a low energy effective field
theory.

Finally, we have advocated the use of anthropic arguments in a
rather different context, that of vacuum selection in M-theory. We
propose to eliminate dynamically stable but phenomenologically
repugnant vacua by showing that they cannot support life.  The aim
is to use only Weinberg's structure formation criterion in making
the decision about whether a given vacuum state is hostile to
life.  We found that a large class of vacua might be eliminated by
such arguments, but that the jury is still out on vacua with exact
$N=1$ SUSY in four dimensions.  The issue may hinge on questions
about the prevalence of weakly coupled gauge theories
and stable particles.
What we view as exciting about
this possibility is that these are questions which might be
amenable to resolution in the not too distant future.

\acknowledgments
The work of T. Banks was supported in part by the Department of
Energy, grant DOE DE-FG02-96ER40559; that of M.D. was supported in part by
the U.S. Department of Energy.  We would like to thank Greg Moore for
collaborating with us on preliminary considerations which led to this
work, as well as for many insightful questions and suggestions.  We would
also like to thank Raphael Bousso for useful conversations. 


\end{document}